\documentclass{ws-mpla}

\begin{document}

\markboth{Ulrich Mosel} {Chiral Symmetry in Nuclei}

\catchline{}{}{}{}{}

\title{CHIRAL SYMMETRY IN NUCLEI\\ THEORETICAL EXPECTATIONS AND HARD FACTS}

\author{\footnotesize ULRICH MOSEL}

\address{Institut f\"ur Theoretische Physik, Universitaet Giessen\\
Giessen, D-35392 , Germany\footnote{mosel@physik.uni-giessen.de}}

\maketitle

\pub{Received (Day Month Year)}{Revised (Day Month Year)}

\begin{abstract}
It is widely believed that chiral symmetry is restored not only at high
temperatures, but also at high nuclear densities. The drop of the order
parameter of the chiral phase transition, the chiral condensate, with
density has indeed been calculated in various models and is as such a
rather robust result. In this talk I point out that the connection of this
property with actual observables is far less clear. For this task a good
hadronic description of the primary production of hadrons, their
propagation inside the nuclear medium, their decay and the propagation of
the decay products through the medium to the detector all have to be
treated with equal accuracy and weight. In this talk I illustrate with the
examples of $\omega$ production and $\pi^0\pi^0$ production how important
in particular final state interactions are.

\end{abstract}

\ccode{21.65.Jk, 25.20.-x, 13.60.Le}

\section{Introduction}

The interest in in-medium properties of hadrons has been growing over the
last decade because of a possible connection with broken symmetries of QCD
and their partial restoration inside nuclear matter. Nearly two decades
ago there were sophisticated theoretical predictions that masses of vector
mesons should generally decrease in medium as a function of density due to
a partial restoration of chiral symmetry \cite{Brown-Rho,HL}.
Specifically, there existed well-worked out predictions that the scalar
$2\pi$ strength should decrease in medium and that the vector meson masses
should drop \cite{Kuni,Klingl}. All of these calculations were performed
for idealized situations (infinite cold nuclear matter at rest) and little
attention was paid to the actual observability of these predicted changes.
At the same time experiments (CERES, TAPS) seemed to show the predicted
behavior. The CERES results indicated a significant broadening of the
$\rho$ meson in medium \cite{Ceres}, whereas the TAPS results on the
$2\pi$ strength exhibited the predicted lowering of the $\sigma$ strength
inside nuclei \cite{TAPS2pi}. Most recently, the CB/TAPS@ELSA experiment
has also obtained an indication for a lowering of the $\omega$ meson mass
in nuclei \cite{Trnka}. Other interesting data in this context have been
obtained by groups at JLAB \cite{g7}, KEK \cite{KEK} and CERN \cite{NA60}.
While experiments with relativistic or ultrarelativistic heavy ions have
the advantage that high densities can be produced, they have the
disadvantage that any observed signal naturally is a time-integral over
the whole collision history. On the other hand, experiments with
microscopic probes on cold nuclei have the advantage that they proceed
closer to equilibrium, even though the densities probed are much lower.

We have focussed our work on in-medium properties of hadrons on two steps.
First, we have performed state-of-the-art calculations of vector meson
spectral functions in cold nuclear matter. Second, we have followed
closely the CB/TAPS@ELSA experiment and have performed various feasibility
studies and analyses of this experiment searching for in-medium changes of
the $\omega$ meson in medium. In a third step we have also analyzed the
lowering of scalar strength in nuclei, predicted as a signature for chiral
symmetry restoration, and indeed observed by the TAPS experiment.

\section{In-Medium Properties of Vector Mesons}

\subsection{QCD sum rules and hadronic models}

While the predicted drop of the chiral condensate with density
\cite{Weise} is a rather robust result  the connection between chiral
condensates on one hand and hadronic spectral functions on the other
cannot simply be inferred from looking at the dependence of the condensate
on density and temperature. Instead, the connection is much more indirect;
only an integral over the spectral function can be linked via the QCD sum
rule to the condensate behavior in medium. In an abbreviated form this
connection is given by
\begin{eqnarray}
R^{{\rm OPE}}(Q^2) & = & {\tilde c_1 \over Q^2} + \tilde c_2 -{Q^2 \over
\pi} \int\limits_0^\infty \!\! ds \, {{\rm Im}R^{{\rm HAD}}(s) \over
(s+Q^2)s}
  \label{eq:opehadr}
\end{eqnarray}
with $Q^2:= -q^2 \gg 0$ and some subtraction constants $\tilde c_i$. Here
$R^{\rm OPE}$ represents a Wilson's operator expansion of the
current-current correlator in terms of quark and gluon degrees of freedom
in the space-like region. It is an expansion in terms of powers of $1/Q^2$
and contains the condensates as expansion parameters. On the other hand,
$R^{{\rm HAD}}(s)$ in (\ref{eq:opehadr}) is the same object for time-like
momenta, represented by a parametrization in terms of hadronic variables.
The dispersion integral connects time- and space-like momenta. Eq.
(\ref{eq:opehadr}) connects the hadronic with the quark world. It allows
to determine parameters in a hadronic parametrization of $R^{{\rm
HAD}}(s)$ by comparing the lhs of this equation with its rhs. Invoking
vector meson dominance it is easy to see that for vector mesons Im$R^{{\rm
HAD}}(s)$ in (\ref{eq:opehadr}) is just the spectral function of the meson
under study.

The operator product expansion of $R^{\rm OPE}$ on the lhs involves quark-
and gluon condensates \cite{Leupold,LeupoldMosel}; of these only the
two-quark condensates are reasonably well known, whereas our knowledge
about already the four-quark condensates is rather sketchy.

Using the measured, known vacuum spectral function for $R^{\rm HAD}$
allows one to obtain information about these condensates appearing on the
lhs of (\ref{eq:opehadr}). On the other hand, modelling the
density-dependence of the condensates yields information on the change of
the hadronic spectral function when the hadron is embedded in nuclear
matter. Since the spectral function appears under an integral the
information obtained is not direct. Therefore, as Leupold et al.\ have
shown \cite{Leupold,LeupoldMosel}, the QCDSR provides important
constraints for the hadronic spectral functions in medium, but it does not
fix them. Kaempfer et al have turned this argument around by pointing out
that measuring an in-medium spectral function of the $\omega$ meson could
help to determine the density dependence of the higher-order condensates
\cite{Kaempfer}. Leupold has also argued that the four-quark condensates
are interesting, because a special one of them can be unambiguously linked
to a difference of spectral functions in the axial and vector channel
\cite{Leupold4q}.

Thus models are needed for the hadronic interactions. The quantitatively
reliable ones can at present be based only on 'classical' hadrons and
their interactions. In lowest order in the density the mass and width of
an interacting hadron in nuclear matter at zero temperature and vector
density $\rho_v$ are given by (for a meson, for example)
\begin{eqnarray}     \label{trho}
{m^*}^2 & =  & m^2 - 4 \pi \mathcal{R}e f_{m N}(q_0,\theta = 0)\, \rho_v
\nonumber \\
m^* \Gamma^* & = & m \Gamma^0 -  4 \pi \mathcal{I}m f_{mN}(q_0,\theta =
0)\, \rho_v ~.
\end{eqnarray}
Here $f_{mN}(q_0,\theta = 0)$ is the forward scattering amplitude for a
meson with energy $q_0$ on a nucleon. The width $\Gamma^0$ denotes the
free decay width of the particle. For the imaginary part this is nothing
other than the classical relation $\Gamma^* - \Gamma^0 = v \sigma \rho_v$
for the collision width, where $\sigma$ is the total cross section. This
can easily be seen by using the optical theorem. The calculation of
hadronic spectral functions in medium thus reduces -- in lowest order in
the density -- to a calculation of hadron-nucleon scattering amplitudes.

Unfortunately it is not a-priori known up to which densities the
low-density expansion (\ref{trho}) is useful. Post et al. \cite{Post1}
have recently investigated this question in a coupled-channel calculation
of selfenergies. Their analysis comprises pions, $\eta$-mesons and
$\rho$-mesons as well as all baryon resonances with a sizeable coupling to
any of these mesons. The authors of \cite{Post1} find that already for
densities less than $0.5 \rho_0$ the linear scaling of the selfenergies
inherent in (\ref{trho}) is badly violated for the $\rho$ and the $\pi$
mesons, whereas it is a reasonable approximation for the $\eta$ meson.
This may serve as a warning sign for many in-medium calculations that use
the low-density approximation without any further proof of its
reliability. Reasons for this deviation from linearity are Fermi-motion,
Pauli-blocking, selfconsistency and short-range correlations. For
different mesons different sources of the discrepancy prevail: for the
$\rho$ and $\eta$ mesons the selfconsistency acts against the low-density
theorem by inducing a strong broadening for the $D_{13}(1520)$ and a
slightly repulsive mass shift for the $S_{11}(1535)$ nucleon resonances to
which the $\rho$ and the $\eta$ meson, respectively, couple. The
investigation of in-medium properties of mesons, for example, thus
involves at the same time the study of in-medium properties of nucleon
resonances and is therefore a coupled-channel problem.

\begin{figure}[htb]
\centerline{\epsfig{file=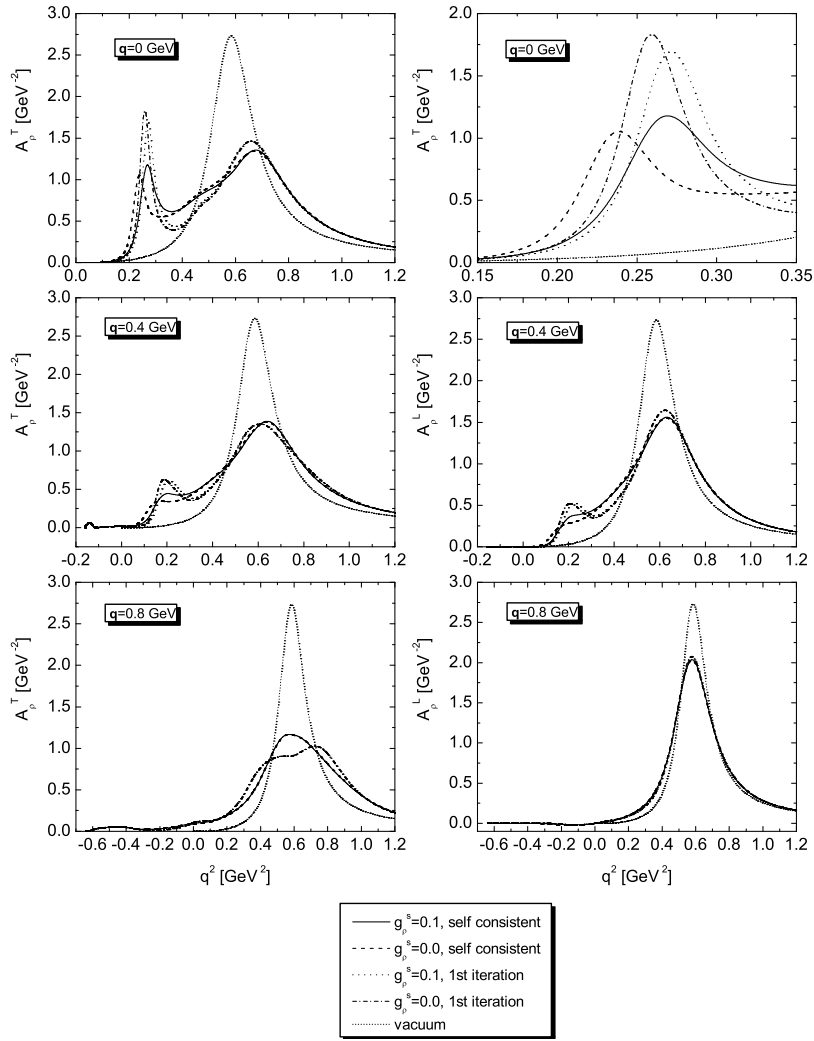,height=14cm}} \caption{Spectral
function of the $\rho$ meson in nuclear matter at density $\rho_0$ for
various momenta indicated in the figure. The left column shows the
transverse spectral function, the right column that of longitudinally
polarized $\rho$ mesons. The thin dotted line in each figure is the vacuum
spectral function, the other curves give the effect of selfconsistency and
short-range correlations (from \protect\cite{Post1}).} \label{rhospect}
\end{figure}
The strong interplay between the $\rho$ meson and the D13(1520)-nucleon
hole excitation leads to a dominant lower hump in the $\rho$ spectral
function also in this relativistic and selfconsistent calculation; it
confirms our earlier result obtained in a more simplified approach.
Whereas the longitudinal component of the $\rho$ meson only broadens
somewhat, the transverse component shows a major distortion which evolves
as a function of the $\rho$ momentum (see Fig. \ref{rhospect}) and
reflects the coupling to the D13(1520) resonance. It is obvious that such
a rich structure cannot be obtained from QCD sumrules that involve only an
integral over the spectral function. At the same time, the D13(1520)
resonance broadens considerably due to the opening of phase-space for
$\rho$-decay. For the $\eta$ meson the optical potential resulting from
our model is rather attractive whereas the in-medium modifications of the
S11(1535) are found to be quite small.

\subsection{Coupled channel calculation of hadronic selfenergies}
Besides the range of validity of the low-density approximation the
meson-nucleon cross sections are often not directly measurable and can be
inferred only from unitarity arguments involving many coupled channels.
Unitarity thus provides important contraints for reactions such as $\omega
N \to 2\pi N$ that are experimentally not accessible, but are not only
needed for but contribute significantly to a calculation of the $\omega$
selfenergy. Bearing this in mind we have recently performed a calculation
of the selfenergy of the $\omega$ meson in medium.
\begin{figure}[htb]
\centerline{\epsfig{file=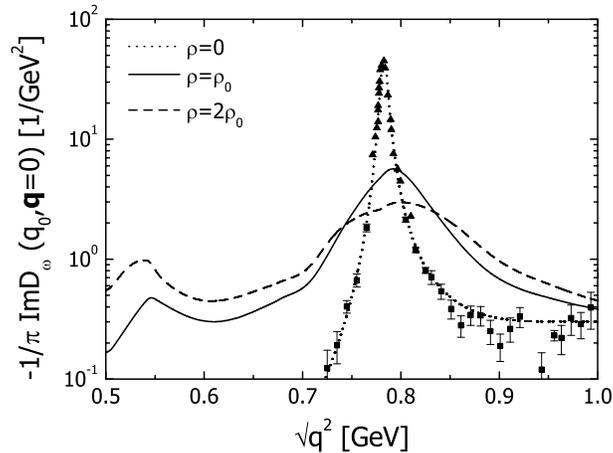,height=6cm}}
 \caption{Spectral function of the $\omega$ meson in nuclear matter at rest,
at densities $0$, $\rho_0$ and 2 $\rho_0$ (from
\protect\cite{Muehlich:2006nn}).} \label{omspectr}
\end{figure}
This calculation is based on a unitary coupled channel analysis of all
existing $\pi N$ and $\gamma N$ data up to an invariant mass of 2 GeV
\cite{Muehlich:2006nn}. The coupled channel character of this calculation
is of utmost importance here because it is the only way to include
experimental constraints on the $2\pi$ decay channel that was found to be
dominant in \cite{Klingl}. This analysis and thus also the selfenergy of
the $\omega$ meson extracted from the $\omega N$ scattering amplitude
gives a broadening of about 60 MeV at $\rho_0$ for an $\omega$ at rest and
a small upward shift of the peak mass. In addition, due to a nonzero
coupling of the $\omega$ to the S11(1535) resonance the $\omega$ spectral
function exhibits a small peak at a mass of around 550 MeV. This
calculation gives, for the first time, the $\omega$ selfenergy also for
nonzero momenta (which corresponds to the experimental situation). The
result of this calculation is shown in Fig. \ref{omspectr}.

For vanishing $\omega$ momentum this result qualitatively agrees with that
of \cite{Lutz} in that it yields a lower mass structure in the spectral
function at an invariant mass of 500 - 600 MeV and only a very small shift
of the main peak; the latter is in contrast to the results of
\cite{Klingl} (for a recent discussion of the results obtained in
\cite{Klingl} see \cite{Eichst}) and of \cite{Thomas}. The latter was
based on a relativistic mean-field model and does not contain any
dispersive effects. Fig.\ \ref{omspectr} also shows clearly that the
$\omega$ meson in medium acquires a significant broadening, caused by
collisions with the surrounding nucleons.

\subsection{Data analysis with GiBUU}
Our second line of approach to the problem of in-medium selfenergies has
concentrated on an analysis of the recent CB/TAPS@ELSA data \cite{Trnka}
Since the experiment looks for the channel $\gamma + A \to A^* + \omega
\to A^* + \gamma + \pi^0$ it is mandatory to control the effects of final
state interactions on the $\pi^0$ in a quantitative way. The only method
available for this is that of coupled channel semiclassical transport
calculations which -- as we had shown earlier in extensive work -- can
give a consistent description of many experimental phenomena, both in
heavy-ion \cite{Lari} as well as in nucleon-, pion- \cite{Buss}, photon-
\cite{Buss-Leitner} and neutrino-induced reactions \cite{Leitner}. For any
reaction on nuclei with hadrons in the final state a state-of-the-art
transport calculation of the final state interactions is an indispensible
part of the theory. We have, therefore, spent significant effort on
developing a new code, dubbed 'GiBUU', for the transport calculations.
This code is written in object-oriented FORTRAN 95/2003 and incorporates
all the experience we have gained with earlier numerical implementations
at Giessen over the last 20 years \cite{GiBUU}. A major effort has gone
into an analysis of the $\omega$ photo-production experiment at
CB/TAPS@ELSA.

\begin{figure}[htb]
\centerline{\epsfig{file=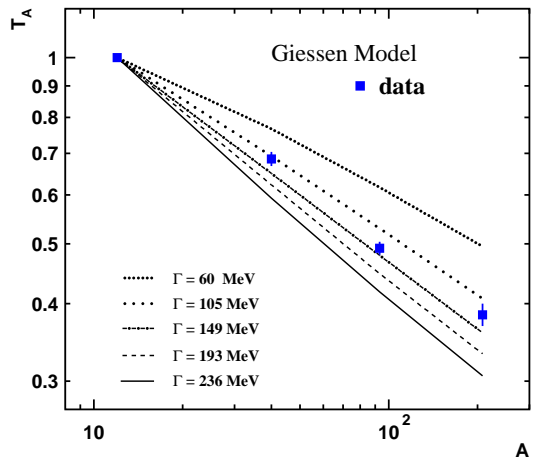,height=8cm}} \caption{Experimentally
determined transparency ratio (squares) as a function of massnumber in
comparison with GiBUU calculations assuming various values for the
inleastic width at an $\omega$ momentum of 1.1 GeV (from
\protect\cite{Metag,Kot}).} \label{TA}
\end{figure}

For $\omega$ mesons the CB/TAPS@ELSA collaboration has measured the
nuclear transparency \cite{Metag,Kot}, see Fig.\ \ref{TA}. This
transparency gives directly the imaginary part of the meson's selfenergy
in medium; using a low-density approximation one can then extract the
inelastic $\omega N$ cross section. The figure shows an analysis of these
data for various values of the collisional width, assumed at a momentum of
1.1 GeV, the average momentum of the $\omega$ mesons in this experiment.
Again, care has been taken to ensure consistency between the collisional
width and the inelastic $\omega N$ cross sections. It is seen from this
figure that a width of about 130 - 150 MeV describes the data. This width
is larger by about a factor of 3 than that previously assumed. A fit to
the data can actually also determine the momentum-dependence of this cross
section \cite{Kot}.
With this method we have also analysed results on the experimental
determination of the nuclear transparency ratio for $\phi$ mesons
\cite{Muhlich:phi}, measured by a group at SPRING8. Also in this case an
unexpectedly large inelastic cross section for $\phi N$ interactions was
extracted. We have found that indeed cross sections about a factor 3
larger than those theoretically expected are needed to explain the
mentioned data, in line with a simple Glauber analysis by the SPRING8
group.

Our simulations give a full event analysis and thus allow to calculate
also background contributions on the same footing as the actual signal.
They also allow insight into the effects of rescattering of the pions
produced in the decay of the $\omega$ meson and have suggested a method to
suppress the rescattered pion background that has actually been adopted by
the experimental group. The result of this analysis is that the $\omega$
spectral function published by the CB/TAPS@ELSA collaboration can be
explained only if a lowering of the $\omega$ meson mass in medium by about
16 \% is assumed together with the appropriate collisional broadening.

A problem in this context is that the experiment does not determine the
spectral function of the $\omega$ meson itself. Instead, we have noted
that the result of the experimental analysis is the product of the primary
production cross section with the spectral function and the partial decay
probability into the channel under study ($\pi^0 \gamma$ here). If the
first and the latter depend strongly on the invariant mass itself, as it
does for the CB/TAPS@ELSA experiment, then significant distortions of the
spectral function may arise. This is a topic that is presently under
intensive study \cite{MosEr}.

\section{In-medium Properties of Scalar Mesons}

Since the restoration of chiral symmetry should lead to a convergence of
scalar-isoscalar and scalar-isovector strength at high densities it was
predicted that the decay of the scalar $\sigma$ meson into two uncharged
pions should become increasingly difficult and the scalar strength should
become narrower at the $2\pi$ threshold \cite{Kuni}. The TAPS
collaboration had indeed initially seen this effect in the $2\pi^0$ data
as predicted, whereas a comparison measurement in the $\pi^0 \pi^{\pm}$
channel did not show such an effect \cite{TAPS2pi}. Various explanations
for these findings have been advanced by the Valencia group \cite{Oset2pi}
and by a group in Lyon \cite{Chanfray} in terms of $\pi - \pi$
correlations or chiral symmetry restoration in nuclei.

\begin{figure}[htb]
\centerline{\epsfig{file=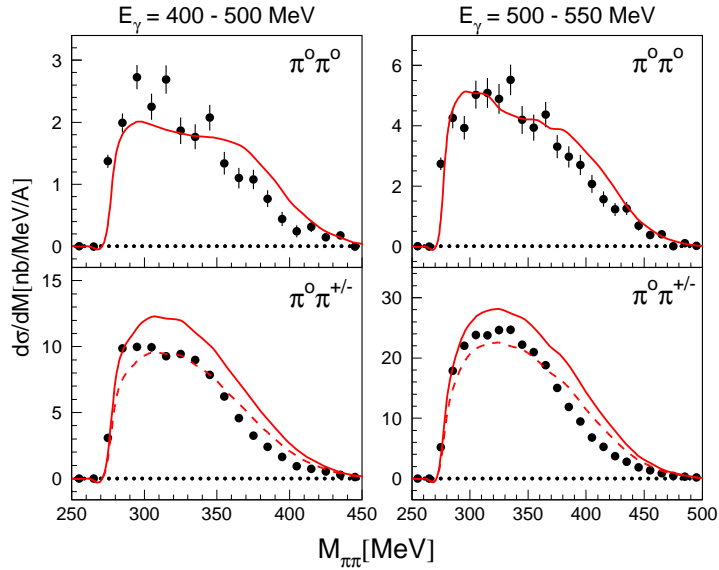,height=8cm}} \caption{Data of the TAPS
collaboration (Bloch et al.) for 2$\pi^0$ photoproduction on a $^{40}Ca$
target for two different photon energies. The solid curve gives the result
of a GiBUU calculation \protect\cite{Buss1}, the dashed curves in the
semicharged 2$\pi$ channel are normalized to the data (see text). Data
from \protect\cite{Bloch}.} \label{2pi}
\end{figure}

None of these calculations, however, did look into the simplest possible
explanation of the observed effects in terms of mundane pion rescattering.
We have, therefore, performed such calculations \cite{Buss} using the
GiBUU method which is ideally suited for this task. These calculations,
which did not contain any effects connected with $\pi \pi$ correlations,
did reproduce the observed effect for the $\pi^0 \pi^0$ channel, but they
also predicted a similar effect in the semi-charged channel where it had
not been seen experimentally. However, a more recent analysis with higher
statistics by Bloch et al. \cite{Bloch}  yielded a result for both charge
channels that is in perfect agreement with our calculations (see Fig.
\ref{2pi}; the dashed lines in this figure are normalized in height to the
data, this normalization reflects uncertainties in the elementary cross
sections). In particular the yield in the semi-charged channel is strongly
influenced by a coupled-channel effect, the charge transfer in $\pi N$
interactions; Glauber based absorption models miss this contribution. This
illustrates that a very sophisticated treatment of final state
interactions is absolutely mandatory when looking for more 'exotic'
effects in nuclei. We conclude that any analysis of the $2\pi^0$ data with
respect to a lowering of the scalar strength in nuclei has to take the
pion rescattering effects into account. Present day's data are all
consistent with simple rescattering.

\section{Summary}
The main message we have learned from the studies reported here is that it
is important to calculate not only in-medium properties under idealized
conditions (static, uniform matter in equilibrium), but to also explore
the influence of these properties on actual observables. The spectral
function itself, which contains the information on in-medium selfenergies,
in particular in-medium masses and widths, is not directly observable.
Instead, both the creation of the studied hadron as well as its decay
influence the observables as much as the spectral function itself and thus
have to be under good control. The same holds for the final state
interactions on hadronic decay products. Here a state-of-the-art treatment
of final state interactions is mandatory. There is now general agreement
on the amount of collisional broadening of vector mesons in medium, but
the verification of an actual mass-shift still requires more work, both
theoretical and experimental. Until then the quest for chiral symmetry
restoration in nuclei remains a challenging topic.

\section{Acknowledgement}
This talk is mainly based on results obtained by Oliver Buss and Pascal
Muehlich as part of their doctoral theses. I thank them, and the entire
GiBUU group, for many inspiring discussions.

\end{document}